\pdfoutput=1
\documentclass[a4paper]{article}
\usepackage{fullpage}

\usepackage{CJKutf8} 

\usepackage{tikz}
\usetikzlibrary{math}
\usetikzlibrary{decorations.pathreplacing}
\usepackage{ifthen}
\usepackage{animate}

\usepackage{subcaption} 
\usepackage{amsmath} 
\usepackage{hyperref}

\newcommand\quixo[4][1]{
	\begin{tikzpicture}[y=-1cm,scale=#1,every node/.style={scale=#1}]
	\draw[ultra thick,rounded corners] (-0.5,-0.5) rectangle (#2-0.5,#2-0.5);
	\foreach \x [count=\xi] in {#3}
		\ifthenelse{\x = 0}{\node[draw, rectangle, rounded corners, fill=black!5, minimum size = 0.98cm] at ({mod(\xi-1,#2)},{int((\xi-1)/#2)}) {};}{
		\ifthenelse{\x = 1}{\node[draw, rectangle, rounded corners, fill=black!5, minimum size = 0.98cm] at ({mod(\xi-1,#2)},{int((\xi-1)/#2)}) {
			\begin{tikzpicture}\draw[thick] (-0.25,0.25) -- (0.25,-0.25) (-0.25,-0.25) -- (0.25,0.25);\end{tikzpicture}
		};}{
		\ifthenelse{\x = 2}{\node[draw, rectangle, rounded corners, fill=black!5, minimum size = 0.98cm, anchor=center] at ({mod(\xi-1,#2)},{int((\xi-1)/#2)}) {
			\begin{tikzpicture}\draw[thick] (0,0) circle (0.25);\end{tikzpicture}
		};}{};};}; 
	#4
	\end{tikzpicture}
}

\newcommand\animatedQuixo[4][1]{
	\begin{tikzpicture}[y=-1cm,scale=#1,every node/.style={scale=#1}]
	\draw[draw=none] (-1.5,-1.5) rectangle (#2+0.5,#2+0.5);
	\ifthenelse{\equal{#4}{}}{}{\node at ({(#2-1)/2},#2) {Move #4};}
	\draw[ultra thick,rounded corners] (-0.5,-0.5) rectangle (#2-0.5,#2-0.5);
	\foreach \x [count=\xi] in {#3} {
		\tikzmath{\a = {int(\x/10)};
						\b = {int(Mod(\x, 10))};
						if \b == 0 then {\dx = 0; \dy=0;} else {
						if \b == 1 then {\dx = -0.5; \dy=0;} else {
						if \b == 2 then {\dx = 0.5; \dy=0;} else {
						if \b == 3 then {\dx = 0; \dy=-0.5;} else {
						if \b == 4 then {\dx = 0; \dy=0.5;} else {
						if \b == 5 then {\dx = -1; \dy=0;} else {
						if \b == 6 then {\dx = 1; \dy=0;} else {
						if \b == 7 then {\dx = 0; \dy=-1;} else {
						if \b == 8 then {\dx = 0; \dy=1;} else {
						};};};};};};};};};
		}
		\ifthenelse{\a = 1}{\node[draw, rectangle, rounded corners, fill=black!5, minimum size = 0.98cm] at ({mod(\xi-1,#2)+\dx},{int((\xi-1)/#2)+\dy}) {
		};}{
		\ifthenelse{\a = 2}{\node[draw, rectangle, rounded corners, fill=black!5, minimum size = 0.98cm] at ({mod(\xi-1,#2)+\dx},{int((\xi-1)/#2)+\dy}) {
			\begin{tikzpicture}\draw[thick] (-0.25,0.25) -- (0.25,-0.25) (-0.25,-0.25) -- (0.25,0.25);\end{tikzpicture}
		};}{
		\ifthenelse{\a = 3}{\node[draw, rectangle, rounded corners, fill=black!5, minimum size = 0.98cm, anchor=center] at ({mod(\xi-1,#2)+\dx},{int((\xi-1)/#2)+\dy}) {
			\begin{tikzpicture}\draw[thick] (0,0) circle (0.25);\end{tikzpicture}
		};}{
		\ifthenelse{\a = 4}{\node[draw, rectangle, rounded corners, fill=red!25, minimum size = 0.98cm] at ({mod(\xi-1,#2)+\dx},{int((\xi-1)/#2)+\dy}) {
		};}{
		\ifthenelse{\a = 5}{\node[draw, rectangle, rounded corners, fill=red!25, minimum size = 0.98cm] at ({mod(\xi-1,#2)+\dx},{int((\xi-1)/#2)+\dy}) {
			\begin{tikzpicture}\draw[thick] (-0.25,0.25) -- (0.25,-0.25) (-0.25,-0.25) -- (0.25,0.25);\end{tikzpicture}
		};}{
		\ifthenelse{\a = 6}{\node[draw, rectangle, rounded corners, fill=blue!25, minimum size = 0.98cm, anchor=center] at ({mod(\xi-1,#2)+\dx},{int((\xi-1)/#2)+\dy}) {
			\begin{tikzpicture}\draw[thick] (0,0) circle (0.25);\end{tikzpicture}
		};}{
		\ifthenelse{\a = 7}{\node[draw, rectangle, rounded corners, fill=blue!25, minimum size = 0.98cm] at ({mod(\xi-1,#2)+\dx},{int((\xi-1)/#2)+\dy}) {
		};};};};};};};}
	}
	\end{tikzpicture}
}

\usepackage[ruled]{algorithm}
\usepackage[noend]{algorithmic}

\title{Quixo Is Solved}
\author{Satoshi Tanaka$^1$ \and François Bonnet$^1$ \and Sébastien Tixeuil$^2$ \and Yasumasa Tamura$^1$}
\date{$^1$ Tokyo Institute of Technology, Tokyo, Japan\\
$^2$ Sorbonne Université, CNRS, LIP6, Paris, France}

\begin{document}

\maketitle

\begin{abstract}
Quixo is a two-player game played on a $5\times5$ grid where the players try to align five identical symbols. Specifics of the game require the usage of novel techniques. Using a combination of value iteration and backward induction, we propose the first complete analysis of the game. We describe memory-efficient data structures and algorithmic optimizations that make the game solvable within reasonable time and space constraints. Our main conclusion is that Quixo is a Draw game. The paper also contains the analysis of smaller boards and presents some interesting states extracted from our computations.
\end{abstract}

\section{Introduction}
\subsection{Quixo}
Quixo is an abstract strategy game designed by Thierry Chapeau in~1995 and published by Gigamic~\cite{bgg-website, gigamic-website}. Quixo won multiple awards, both in United States~\cite{award-usa-mensa, award-usa-100-selection, award-usa-best, award-usa-gold} and in France~\cite{award-france-golden, award-france-oscar}. While a four-player variant exists, Quixo is mostly a two-player game that is played on a $5\times5$ grid, also called \emph{board}. Each grid cell, also called \emph{tile}, can be empty, or marked by the symbol of one player: either X or O.

At each turn, the active player first \emph{(i)} takes a tile -- empty or with her symbol -- from the border (\emph{i.e.} excluding the 9 central tiles), and then \emph{(ii)} inserts it, with her symbol, back into to the grid by pushing existing tiles toward the hole created by the tile removal in step \emph{(i)}. The winning player is the first to create a line of tiles all with her symbol, horizontally, vertically, or diagonally. Note that if a player creates two lines with distinct symbols in a single turn, then the opponent is the winner.

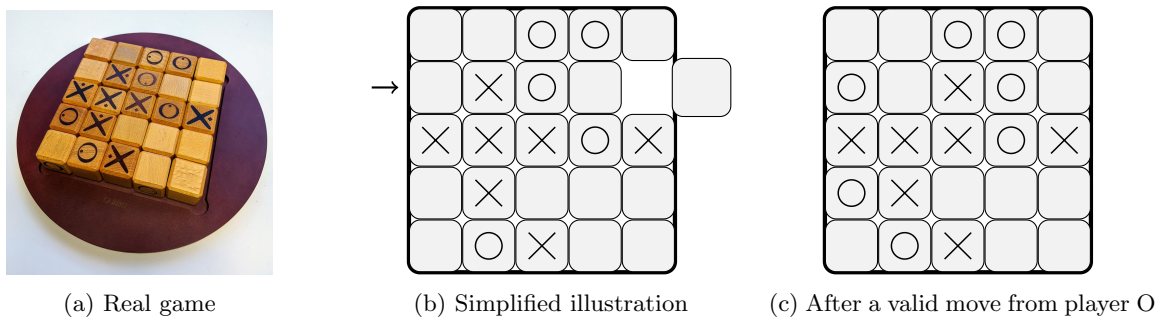
\begin{figure*}[b]
\begin{subfigure}[t]{0.32\textwidth}
    \centering
    \includegraphics[width=3.5cm]{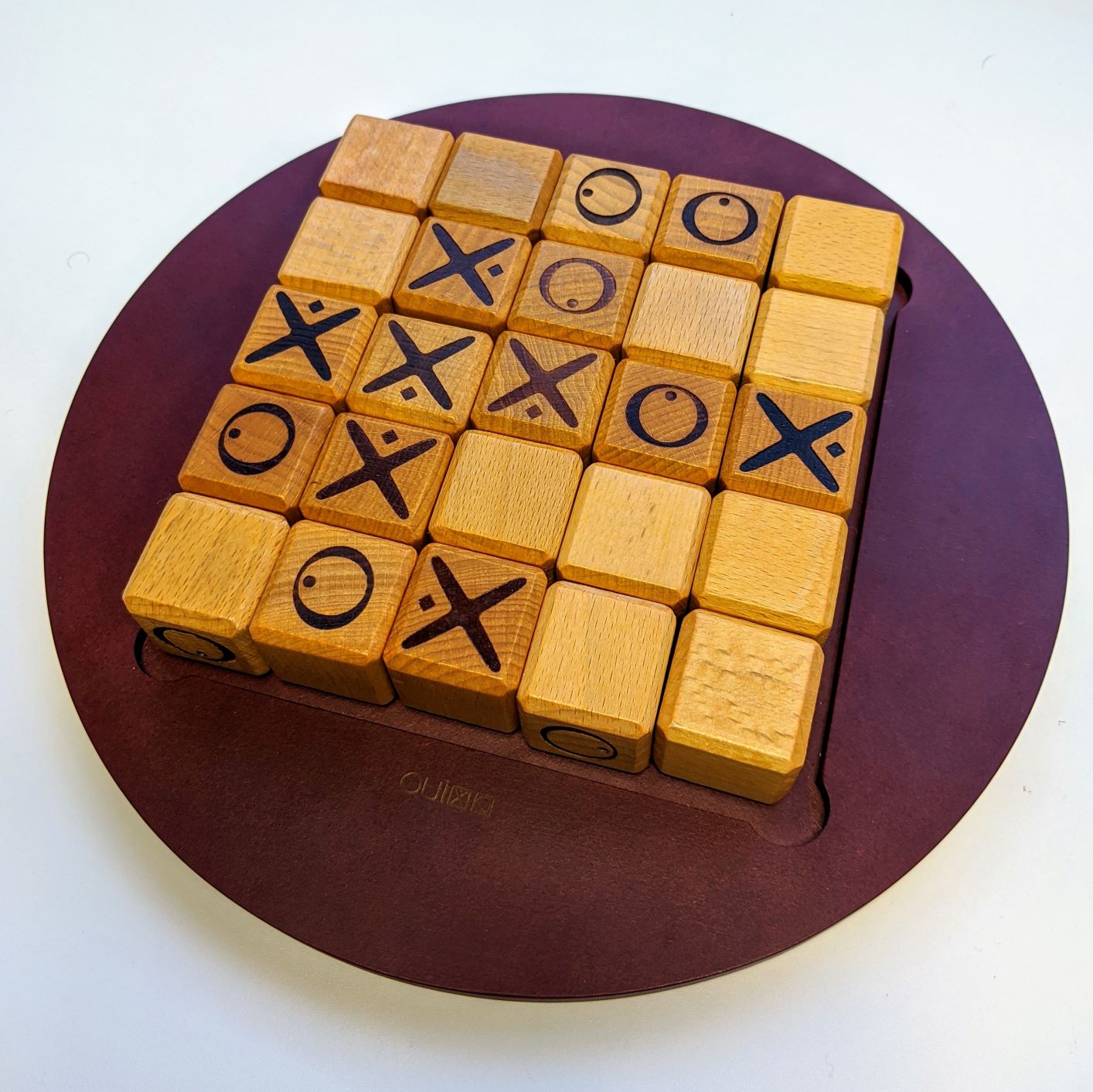}
    \caption{Real game}
    \label{fig:quixo-example-real}
\end{subfigure}\hfill
\begin{subfigure}[t]{0.32\textwidth}
    \centering
    \begin{tikzpicture}[y=-1cm,scale=0.7,every node/.style={scale=0.7}]
	\draw[ultra thick,rounded corners] (-0.5,-0.5) rectangle (4.5,4.5);
	\node[draw, rectangle, rounded corners, fill=black!5, minimum size = 0.98cm] at (0,0) {};
	\node[draw, rectangle, rounded corners, fill=black!5, minimum size = 0.98cm] at (1,0) {};
	\node[draw, rectangle, rounded corners, fill=black!5, minimum size = 0.98cm] at (2,0) {
	    \begin{tikzpicture}\draw[thick] (0,0) circle (0.25);\end{tikzpicture} 
	};
	\node[draw, rectangle, rounded corners, fill=black!5, minimum size = 0.98cm] at (3,0) {
	    \begin{tikzpicture}\draw[thick] (0,0) circle (0.25);\end{tikzpicture} 
	};
	\node[draw, rectangle, rounded corners, fill=black!5, minimum size = 0.98cm] at (4,0) {};
	\node[draw, rectangle, rounded corners, fill=black!5, minimum size = 0.98cm] at (0,1) {};
	\node[draw, rectangle, rounded corners, fill=black!5, minimum size = 0.98cm] at (1,1) {
	    \begin{tikzpicture}\draw[thick] (-0.25,0.25) -- (0.25,-0.25) (-0.25,-0.25) -- (0.25,0.25);\end{tikzpicture} 
	};
	\node[draw, rectangle, rounded corners, fill=black!5, minimum size = 0.98cm] at (2,1) {
	    \begin{tikzpicture}\draw[thick] (0,0) circle (0.25);\end{tikzpicture} 
	};
	\node[draw, rectangle, rounded corners, fill=black!5, minimum size = 0.98cm] at (3,1) {};
	\node[draw, rectangle, rounded corners, fill=black!5, minimum size = 1.1cm] at (4+1,1) {}; 
	\node[draw, rectangle, rounded corners, fill=black!5, minimum size = 0.98cm] at (0,2) {
	    \begin{tikzpicture}\draw[thick] (-0.25,0.25) -- (0.25,-0.25) (-0.25,-0.25) -- (0.25,0.25);\end{tikzpicture} 
	};
	\node[draw, rectangle, rounded corners, fill=black!5, minimum size = 0.98cm] at (1,2) {
	    \begin{tikzpicture}\draw[thick] (-0.25,0.25) -- (0.25,-0.25) (-0.25,-0.25) -- (0.25,0.25);\end{tikzpicture} 
	};
	\node[draw, rectangle, rounded corners, fill=black!5, minimum size = 0.98cm] at (2,2) {
	    \begin{tikzpicture}\draw[thick] (-0.25,0.25) -- (0.25,-0.25) (-0.25,-0.25) -- (0.25,0.25);\end{tikzpicture} 
	};
	\node[draw, rectangle, rounded corners, fill=black!5, minimum size = 0.98cm] at (3,2) {
	    \begin{tikzpicture}\draw[thick] (0,0) circle (0.25);\end{tikzpicture} 
	};
	\node[draw, rectangle, rounded corners, fill=black!5, minimum size = 0.98cm] at (4,2) {
	    \begin{tikzpicture}\draw[thick] (-0.25,0.25) -- (0.25,-0.25) (-0.25,-0.25) -- (0.25,0.25);\end{tikzpicture} 
	};
	\node[draw, rectangle, rounded corners, fill=black!5, minimum size = 0.98cm] at (0,3) {};
	\node[draw, rectangle, rounded corners, fill=black!5, minimum size = 0.98cm] at (1,3) {
	    \begin{tikzpicture}\draw[thick] (-0.25,0.25) -- (0.25,-0.25) (-0.25,-0.25) -- (0.25,0.25);\end{tikzpicture} 
	};
	\node[draw, rectangle, rounded corners, fill=black!5, minimum size = 0.98cm] at (2,3) {};
	\node[draw, rectangle, rounded corners, fill=black!5, minimum size = 0.98cm] at (3,3) {};
	\node[draw, rectangle, rounded corners, fill=black!5, minimum size = 0.98cm] at (4,3) {};
	\node[draw, rectangle, rounded corners, fill=black!5, minimum size = 0.98cm] at (0,4) {};
	\node[draw, rectangle, rounded corners, fill=black!5, minimum size = 0.98cm] at (1,4) {
	    \begin{tikzpicture}\draw[thick] (0,0) circle (0.25);\end{tikzpicture} 
	};
	\node[draw, rectangle, rounded corners, fill=black!5, minimum size = 0.98cm] at (2,4) {
	    \begin{tikzpicture}\draw[thick] (-0.25,0.25) -- (0.25,-0.25) (-0.25,-0.25) -- (0.25,0.25);\end{tikzpicture} 
	};
	\node[draw, rectangle, rounded corners, fill=black!5, minimum size = 0.98cm] at (3,4) {};
	\node[draw, rectangle, rounded corners, fill=black!5, minimum size = 0.98cm] at (4,4) {};
	
	\draw [->, thick] (-1.2,1)--(-0.7,1); 

	\end{tikzpicture}
    \caption{Simplified illustration}
    \label{fig:quixo-example-tikz}
\end{subfigure}\hfill
\begin{subfigure}[t]{0.32\textwidth}
    \centering
    \quixo[0.7]{5}{0,0,2,2,0, 2,0,1,2,0, 1,1,1,2,1, 2,1,0,0,0, 0,2,1,0,0}{}
    \caption{After a valid move from player O}
    \label{fig:quixo-example-after-move}
\end{subfigure}
\caption{The two-player game of Quixo}
\label{fig:quixo-example}
\end{figure*}

Figures~\ref{fig:quixo-example-real} and~\ref{fig:quixo-example-tikz} show the real game and our corresponding representation. Figure~\ref{fig:quixo-example-after-move} depicts the resulting board after a valid turn by player O from the board depicted in Figure~\ref{fig:quixo-example-tikz}: player O first \emph{(i)} takes the rightmost (empty) tile of the second row, and then \emph{(ii)} inserts it at the leftmost position shifting the other tiles of this second row to the right. A complete game on a smaller $4\times4$ board is given in Figure~\ref{fig:4x4_optimal_play} on page~\pageref{fig:4x4_optimal_play}.

Quixo bears an immediate resemblance with some classical games such as Tic-Tac-Toe, Connect-Four, or Gomoku. However there are two major differences: \emph{(1)} the board is ``dynamic''; a placed X or O may change its location in subsequent turns, \emph{(2)} the game is unbounded (in term of turns). The first point is what makes the game interesting to play: dynamicity makes Quixo very difficult for a human player to plan more than a couples of turns in advance. The second point raises immediately a natural scientific question about termination. Trivially, players could ``cooperate'' to create an infinite game, as official rules do not specify any terminating rules, such as the 50-move or the threefold repetition rules of chess. For example, by always taking the same tiles and having only one X and one O on the board. However, it remains unclear whether an infinite game exists when both players play to win.

\subsection{Related work}
From the seminal work on Nim~\cite{bouton1901nim} to more recent results on Poker~\cite{Bowling2015} or Go~\cite{Silver2016}, studying games has always been a prolific research topic. Considered as recreational mathematics, results were initially obtained by theoretical analysis and/or handmade computations. Later, in order to solve more complex games, it became necessary to use computer-assisted techniques. However, even for complex games, the outcome is sometimes known without any computations. Hex was proven by Nash to be a First-Player-Win game using a now-classic strategy-stealing argument~\cite{gardner59}. Yet, it is still extremely complex to find optimal strategies; larger boards (up to $10\times10$) are regularly solved~\cite{Pawlewicz2013} but a winning strategy for the original $11\times11$ is still missing.

In 1988, Allen and Allis independently proved that Connect-Four is a First-Player-Win game~\cite{allis1988knowledge}. Awari and Checkers were both proven to be Draw games, respectively by Romein and Bal in 2003~\cite{romein2003solving}, and by Schaeffer et al. in 2007~\cite{schaeffer2007checkers}. Among connection games, it is worth mentioning that Gomoku and Renju have also been proved to be First-Player-Win games~\cite{Allis1996, wagner2001solving}.

Note that aforementioned results about Go and Poker are quite different. These works propose excellent strategies, well-above human capacities, but they do not consist in a proper solution of either game. The exact outcome and optimal strategies remain unknown. In contrast, here, we are interested in solving Quixo with the usual meaning of obtaining exact results, similar to Connect-Four, Awari, Checkers.


\subsection{Objectives and challenges}\label{sec:goals}

Our main goal is to solve Quixo, which means finding the optimal outcome of the game assuming perfect players. As for any combinatorial game, there are only three possible outcomes:
first-player-Win, second-player-Win, or Draw. While a second-player-Win seems unlikely, there is no easy observation that would permit to discard it.\footnote{A typical strategy-stealing argument cannot be applied, at least not in an obvious way.} 
It is improbable to obtain analytical results, so we focus on computing this optimal outcome and the corresponding optimal strategies. More precisely, we are looking for outcomes of all states, not only from the initial state. In game terminology, this is called strongly-solving the game. Finally, in addition to the real $5\times5$ game of Quixo, we also analyze a variant using $4\times4$ grid.

Even on the $5\times5$ grid, Quixo's game ``tree'' is not extremely large, with respect to other games. The number of positions is upper bounded by $2\cdot3^{25}\approx 1.7\cdot10^{12}$ configurations -- 2 possibilities for the active player, and 3~options for each cell in the grid. These numbers are in a similar order of magnitude as the numbers of positions in Connect-Four, which was solved 30 years ago~\cite{allis1988knowledge}.

However, the game ``tree'' of Quixo is very different from other similar game ``trees''. For most games, the ``trees'' are directed acyclic graphs (DAG), assuming the merging of identical positions reached from different histories. Conversely, for Quixo, the game ``tree'' contains cycles (again, assuming the same kind of position merging). Indeed, especially in the late game, when the grid is mostly full of Xs and Ox, most moves do not add symbols, only reorganize them. Therefore, a simzople minimax algorithm (with or without alpha-beta pruning) may never terminate.

As a consequence, instead of searching the game ``tree'' with a DFS algorithm (as done by minimax or alike algorithms), it is necessary to use a more costly approach. Ideally, one would like to analyze the whole game ``tree'', but it is currently impossible to store it all at once in memory on commodity hardware.

\subsection{Results overview}
Our solution involves a combination of backward-induction and value-iteration algorithms, implemented using a state representation that is both time and space efficient. Based on our computations, the regular $5\times5$ Quixo is a Draw. In more details, neither player has a winning strategy if both players play optimally, and the game continues forever. On smaller grids, the first player wins. Interestingly, on the $4\times4$ grid, it takes at least 21 moves (11~moves from the first player and 10~from the opponent) to win. Since $21>16$, it is always necessary to re-use existing tiles.

\paragraph{Outline.}
Section~\ref{sec:prelim} presents some basic definitions and terminology used in the paper. Sections~\ref{sec:solving} and~\ref{sec:algo} describe respectively the data structures and the algorithms used to solve Quixo. Section~\ref{sec:result} summarizes our findings and highlights some unexpected observations. It also includes a nice animation of an optimal game run. Finally, section~\ref{sec:conclusion} concludes the paper with a list of open problems.

\section{Preliminaries}\label{sec:prelim}
By convention, the \emph{first player} is player~X and the \emph{second player} is player~O. The \emph{board} corresponds to the 25 tiles and the \emph{active player} denotes the player playing next. Note that, contrarily to TTT or Connect4, the active player cannot be deduced automatically from a given board. Therefore a \emph{state} of the game consists of a board and an active player. The \emph{initial state} is the empty board (that is, the board with 25 empty tiles) with player~X as active player.

A state is \emph{terminal} if its board contains a line of Xs or Os tiles. The \emph{children} of a given state are all states obtained by a move of the active player. A terminal state has no children since the game is over and there is no valid moves. The \emph{parents} of a state are defined analogously. The set of states and parent-child relations induce the \emph{game graph} of Quixo (referred to as the game ``tree'' in the prequel). As mentioned earlier, this graph is neither a tree nor an acyclic graph.

\paragraph{Outcomes.}
Each state has a \emph{state value}, also called \emph{outcome} which can be either \emph{active-player-Win}, \emph{active-player-Loss}, or \emph{Draw}. For brevity, the \emph{active-player} part is omitted, and a Win (resp. Loss and Draw) state denotes a state whose outcome is Win (resp. Loss and Draw). The outcome of a terminal state is trivially defined. For non-terminal states, the outcome is inductively defined as follows:
\begin{itemize}
    \item Win if there is at least one Loss child,
    \item Loss if all children are Win,
    \item Draw otherwise.
\end{itemize}

\begin{figure*}[b]
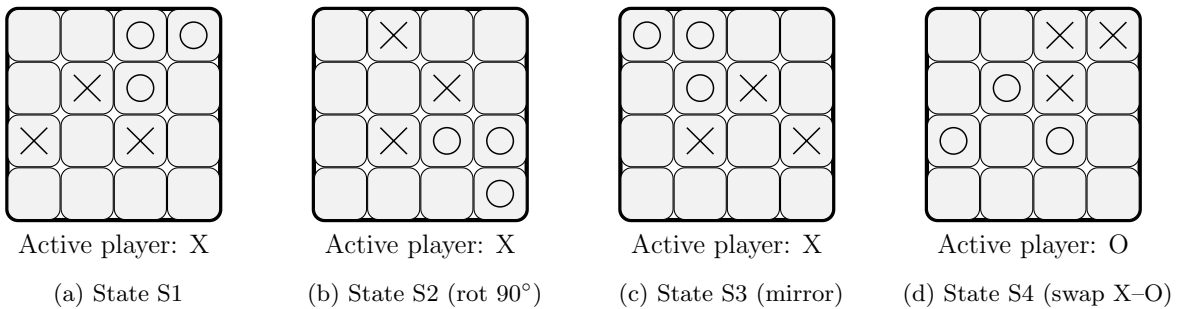

\begin{subfigure}[t]{0.24\textwidth}
    \centering
    \quixo[0.7]{4}{0,0,2,2, 0,1,2,0, 1,0,1,0, 0,0,0,0}{\node at (1.5,4) {\Large Active player: X};}
    \caption{State S1}
\end{subfigure}\hfill
\begin{subfigure}[t]{0.24\textwidth}
    \centering
    \quixo[0.7]{4}{0,1,0,0, 0,0,1,0, 0,1,2,2, 0,0,0,2}{\node at (1.5,4) {\Large Active player: X};}
    \caption{State S2 (rot $90^{\circ}$)}
\end{subfigure}\hfill
\begin{subfigure}[t]{0.24\textwidth}
    \centering
    \quixo[0.7]{4}{2,2,0,0, 0,2,1,0, 0,1,0,1, 0,0,0,0}{\node at (1.5,4) {\Large Active player: X};}
    \caption{State S3 (mirror)}
\end{subfigure}\hfill
\begin{subfigure}[t]{0.24\textwidth}
    \centering
    \quixo[0.7]{4}{0,0,1,1, 0,2,1,0, 2,0,2,0, 0,0,0,0}{\node at (1.5,4) {\Large Active player: O};}
    \caption{State S4 (swap X--O)}
\end{subfigure}
\caption{Equivalent states with respect to symmetries and swapping}
\label{fig:equiv-states}
\end{figure*}

\paragraph{Symmetries and swapping.}
Rotating or mirroring the board does not change the state value. Therefore states can be grouped in equivalence classes. Said differently, states with symmetrical boards can be merged into a single node in the game graph. This optimization divides approximately by eight the number of states: four being due to rotations, and two to vertical mirroring. Note that the horizontal mirroring is the same as vertical mirroring and $180^{\circ}$ rotation. Also, swapping the active player and flipping all Xs and Os to Os and Xs respectively creates a new equivalent state. Figure~\ref{fig:equiv-states} illustrates these notions. All four states are equivalent.

In the remaining of the paper, all states have X as active player. By an abuse of notation, we then identify the state and its board, omitting the active player.

\section{Solving Quixo -- data structures}\label{sec:solving}
This section considers only $5\times5$ Quixo but explanations can easily be adapted for smaller grids.\footnote{Note that larger grids cannot be solved using our state representation: 128bits variables would be necessary.} First we describe our memory efficient representation of states in memory. Then we focus on the more general problem of storing intermediate results.

\subsection{Optimized state representation}\label{sec:state}
\paragraph{Naive approaches.}
The most natural representation of a state in memory is probably to use a 1-dimensional (or 2-dimensional) array of 25 elements where each entry takes values from the set $\{0,1,2\}$ to map $\{\text{empty}, \text{X}, \text{O}\}$. Assuming a typical 1-byte char element, such a representation requires 25~bytes per state.

Of course, using a whole byte to store only 3 values wastes space. Instead, the sequence of 25 numbers could be seen as a single number written in ternary basis. Hence each state corresponds to a unique integer in $\{0, \ldots, 3^{25}-1\}$. Since $\log_2(3^{25})\approx40$, a single 8-byte variable is sufficient to store a state. With respect to memory space, this representation is optimal. However, all basic operations on states are costly: checking if there is a line of Xs or Os, executing a move, etc.

\paragraph{Our optimized approach.}
Still using only 8~bytes, we represent a state as depicted on Figure~\ref{fig:state-64bit}. The 25~first least-significant bits indicate the location of Os on the board. After skipping 7~bits, the 25~following bits indicate the location of Xs. For example, the state of Figure~\ref{fig:quixo-example-tikz} is represented in Figure~\ref{fig:state-64bit-ex}.

\begin{figure*}[ht]
\centering
\begin{tikzpicture}
\foreach \x in {0,...,63} \node at (\x*0.2,0) {0};
\draw (-0.1,-0.25) rectangle (6*0.2+0.1,0.25);
\draw (-0.1,-0.25) -- (6*0.2+0.1,0.25);
\draw [decorate,decoration={brace,amplitude=2mm,mirror,raise=3.5mm}]   (-0.1,0) -- (6*0.2+0.1,0) node[midway,yshift=-8mm]{unused};
\draw (32*0.20-0.1,-0.25) rectangle (38*0.2+0.1,0.25);
\draw (32*0.20-0.1,-0.25) -- (38*0.2+0.1,0.25);
\draw [decorate,decoration={brace,amplitude=2mm,mirror,raise=3.5mm}]   (32*0.2-0.1,0) -- (38*0.2+0.1,0) node[midway,yshift=-8mm]{unused};

\draw (7*0.2-0.1,-0.25) rectangle (31*0.2+0.1,0.25);
\draw [decorate,decoration={brace,amplitude=2mm,raise=3.5mm}]   (7*0.2-0.1,0) -- (31*0.2+0.1,0) node[midway,yshift=8mm]{25bits to store X locations};
\draw (39*0.2-0.1,-0.25) rectangle (63*0.2+0.1,0.25);
\draw [decorate,decoration={brace,amplitude=2mm,raise=3.5mm}]   (39*0.2-0.1,0) -- (63*0.2+0.1,0) node[midway,yshift=8mm]{25bits to store O locations};
\end{tikzpicture}
\caption{State representation in 64bits (LSB on the right)}
\label{fig:state-64bit}
\end{figure*}

\begin{figure*}[ht]
\centering
\begin{tikzpicture}
\draw (-0.1,-0.25) rectangle (6*0.2+0.1,0.25);
\draw (-0.1,-0.25) -- (6*0.2+0.1,0.25);
\draw (32*0.20-0.1,-0.25) rectangle (38*0.2+0.1,0.25);
\draw (32*0.20-0.1,-0.25) -- (38*0.2+0.1,0.25);
\draw (7*0.2-0.1,-0.25) rectangle (31*0.2+0.1,0.25);
\draw (39*0.2-0.1,-0.25) rectangle (63*0.2+0.1,0.25);
\node[anchor=east] at (0,0) {\texttt{s\,=\,}};
\foreach \x [count=\xi] in {0,0,0,0,0,0,0, 0,0,0,0,0, 0,1,0,0,0, 1,1,1,0,1, 0,1,0,0,0, 0,0,1,0,0, 0,0,0,0,0,0,0, 0,0,1,1,0, 0,0,1,0,0, 0,0,0,1,0, 1,0,0,0,0, 0,1,0,0,0} \node at ({(\xi-1)*0.2},0) {\x};

\node[anchor=east] at (0,-0.8) {\texttt{B\,=\,}};
\foreach \x [count=\xi] in {0,0,0,0,0,0,0, 0,0,0,0,0, 1,1,1,1,1, 0,0,0,0,0, 0,0,0,0,0, 0,0,0,0,0, 0,0,0,0,0,0,0, 0,0,0,0,0, 1,1,1,1,1, 0,0,0,0,0, 0,0,0,0,0, 0,0,0,0,0} \node at ({(\xi-1)*0.2},-0.8) {\x};

\node[anchor=east] at (0,-1.6) {\texttt{C\,=\,}};
\foreach \x [count=\xi] in {0,0,0,0,0,0,0, 0,0,0,0,0, 0,0,0,0,0, 0,0,0,0,0, 0,0,0,0,0, 0,0,0,0,0, 0,0,0,0,0,0,0, 0,0,0,0,0, 1,0,0,0,0, 0,0,0,0,0, 0,0,0,0,0, 0,0,0,0,0} \node at ({(\xi-1)*0.2},-1.6) {\x};
\end{tikzpicture}
\caption{Our 64bits representation of the state of Figure~\ref{fig:quixo-example-tikz} and constants \texttt{B} and \texttt{C} used to compute the move creating the state of Figure~\ref{fig:quixo-example-after-move} using \texttt{(((s \& B) >> 1) \& B) | (s \& \textasciitilde B) | C}}
\label{fig:state-64bit-ex}
\end{figure*}

This representation offers some decisive advantages. It enables very fast computation of all basic operations. Given a state \verb#s#, and some appropriate pre-computed constants (see Figure~\ref{fig:state-64bit-ex}), all the following operations can be done efficiently (assuming \texttt{<<}, \texttt{>>}, \texttt{\&}, \verb#|#, and \verb#~# denote the classical bitwise ``left shift'', ``right shift'', ``and'', ``or'', and ``not'', respectively):
\begin{itemize}
\item Swapping the players: \verb# s << 32 | s >> 32#
\item Checking the existence of a tile at a given location:\\
\verb#s & A != 0#, with a different \verb#A# for each pair of tile and symbol.
\item Checking the existence of a given line of Xs or Os:\\
\verb#s & A == A#, with a different \verb#A# for each pair of  line and symbol.
\item More interestingly, moves can also be computed quickly (with different \verb#B# and \verb#C# for each move):

\begin{tabular}{l@{}l}
Left-pushing move: & \verb# (((s & B) << 1) & B) | (s & ~B) | C#\\
Right-pushing move: & \verb# (((s & B) >> 1) & B) | (s & ~B) | C#\\
Down-pushing move: & \verb# (((s & B) >> 5) & B) | (s & ~B) | C#\\
Up-pushing move: & \verb# (((s & B) << 5) & B) | (s & ~B) | C#,
\end{tabular}

The formulas are valid both when considering a marked tile or an empty tile. Moreover, it also allows computing the previous states.
\end{itemize}

Unfortunately, rotations and symmetries are still costly to compute. In fact, we believe that there is no efficient way to compute rotations with a compact data structure. Based on our observations, it is faster to avoid symmetry optimizations, and simply compute independently values for all symmetrical states. In the next section, we thus investigate how to store the outcomes of $3^{25}$ states.

\subsection{Optimized storage of results}\label{sec:storage}

Using our optimized state representation, computations can be done quickly. It remains to consider the problem of storing the outcome of each state. Indeed, in order to strongly-solve the game, we need to record the outcome of all possible states. Three possible outcomes (Win/Loss/Draw) means that 2~bits are necessary to store each outcome. Using a typical (state:\,value) associative array requires at least 64~bits + 2~bits per entry, which sums up to more than 6.5TB.\footnote{In practise, this amount of space is likely much higher to to memory alignment. Furthermore, typical data structure would likely use 128~bits per entry, and most likely a substantial overhead (\emph{e.g.} pointers). So, a more realistic estimation for full storage is in the order of 15TB.}

As mentioned earlier, it is possible to enumerate all possible states in a pre-determined order. It is therefore natural to only store the outcomes in a (giant) bit array. Again, 2~bits per entry yields a total size of $2\cdot3^{25}=197$GB. 
Although more reasonable, renting a server with 200~GB of RAM may still require a significant investment. We further reduce memory requirements. 

The obvious solution is to avoid storing all outcomes at the same time in RAM. Using backward induction (see Section~\ref{sec:algo}), we only need to have a subset of already computed values to compute the new outcomes. For example, to compute all states containing 10~Xs and 8~Os with 8~Xs and 10~Os, it is sufficient to know the (inductively computed) outcomes of states containing either 8~Xs and 11~Os, and 10~Xs and 9~Os.\footnote{Note that this is a simplified explanation. Indeed, as mentioned in Section~\ref{sec:prelim}, we consider only states with X as active player. In our real computations, we need to swap players after each move, so children of states containing 10~Xs and 8~Os may have 8~Xs and 10 or 11~Os.} Therefore we partition the $3^{25}$ states based on the number of Xs and Os. Let ${\cal C}_{x,o}$ denotes the class of states containing $x$~Xs and $o$~Os.

The largest class is ${\cal C}_{8,8}$ which contains ${25\choose 8}\cdot{17\choose 8}\approx2.6\cdot10^{10}$ states. Using 2~bits per state, it corresponds to $\approx6.1$GB of RAM. Since we can implement our algorithm using at most two classes loaded in memory at once, it becomes possible to solve the game on a more typical 16GB-RAM computer. This partitioning seems easy but there is an hidden problem:
\begin{itemize}
\item Creating a bijection between the set of all $3^{25}$ states and the set of natural numbers $\{0,\ldots,3^{25}-1\}$ is straightforward and ``fast enough'' to compute (in both directions). As explained earlier, one can see the 25 cells as the 25-digit ternary representation of a number (0 for empty, 1 for X, and 2 for O).
\item Creating a bijection between ${\cal C}_{x,o}$ and the set of natural numbers $\left\{0,\ldots,{25\choose x}\cdot{25-x\choose o}-1\right\}$ is less straightforward and more difficult to implement, especially in an efficient way. We further describe our scheme in the remaining of the section. 
\end{itemize}

\paragraph{Bijections computation.}
\begin{figure*}[t]
\centering
\quixo[0.7]{5}{0,1,2,1,0, 0,1,1,0,0, 2,2,0,1,1, 0,2,0,0,1, 1,2,0,0,0}
~~\\~~\\

\begin{tikzpicture}
\draw (-0.1,-0.25) rectangle (6*0.2+0.1,0.25);
\draw (-0.1,-0.25) -- (6*0.2+0.1,0.25);
\draw (32*0.20-0.1,-0.25) rectangle (38*0.2+0.1,0.25);
\draw (32*0.20-0.1,-0.25) -- (38*0.2+0.1,0.25);
\draw (7*0.2-0.1,-0.25) rectangle (31*0.2+0.1,0.25);
\draw (39*0.2-0.1,-0.25) rectangle (63*0.2+0.1,0.25);
\foreach \x [count=\xi] in {0,0,0,0,0,0,0, 0,1,0,1,0, 0,1,1,0,0, 0,0,0,1,1, 0,0,0,0,1, 1,0,0,0,0, 0,0,0,0,0,0,0, 0,0,1,0,0, 0,0,0,0,0, 1,1,0,0,0, 0,1,0,0,0, 0,1,0,0,0} \node at ({(\xi-1)*0.2},0) {\x};

\node[anchor=east] at (6*0.2,-0.5) {X:};
\foreach \x [count=\xi] in {0,1,0,1,0, 0,1,1,0,0, 0,0,0,1,1, 0,0,0,0,1, 1,0,0,0,0} \node at ({(\xi-1+7)*0.2},-0.5) {\x};
\node[anchor=west] at (32*0.2,-0.5) {$= S_X$};

\node[anchor=east] at (6*0.2,-1) {O:};
\foreach \x [count=\xi] in {0,0,1,0,0, 0,0,0,0,0, 1,1,0,0,0, 0,1,0,0,0, 0,1,0,0,0} \node at ({(\xi-1+7)*0.2},-1) {\x};

\node[anchor=east] at (6*0.2,-1.5) {filtered O:};
\foreach \x [count=\xi] in {0, ,1, ,0, 0, , ,0,0, 1,1,0, , , 0,1,0,0, ,  ,1,0,0,0} \node at ({(\xi-1+7)*0.2},-1.5) {\x};

\node[anchor=east] at (6*0.2,-2) {shifted O:};
\foreach \x [count=\xi] in {0,0,0,0,0,0,0,0,    0,  1  ,0, 0,    0,0, 1,1,0,     0,1,0,0,     1,0,0,0} \node at ({(\xi-1+7)*0.2},-2)  {\x};
\node[anchor=west] at (32*0.2,-2) {$= S_O$};
\end{tikzpicture}
\caption{A state in the class (8,5), its 64bits representation, and the steps to compute its index}
\label{fig:enc-state}
\end{figure*}

Let us explain with an example the bijection we use in our implementation. Consider the state depicted in Figure~\ref{fig:enc-state} and its state representation (as defined earlier). Let us split the 64-bit variable into two 32-bit variables. For the variable representing the location of the Os, we remove the digits where Xs are located, and then shift remaining digits to the right (see example). In the class ${\cal C}_{8,5}$, the index of this state is then

\begin{equation}
ord(S_X)\cdot{25-8 \choose 5} + ord(S_O),
\label{eq:index}
\end{equation}
where $ord(\cdot)$ is an order function that returns the order of a number among numbers with the same population count (aka. Hamming weight). For simplicity, the population count is returned by function $pop(\cdot)$. 

For a given class ${\cal C}_{x,o}$, there are ${25\choose x} = ord(\underbrace{1\ldots1}_{x}\underbrace{0\ldots0}_{25-x})+1$ valid positions of the Xs in the grid. After placing the Xs, there remain ${25-x\choose o} = ord(\underbrace{0\ldots0}_{x}\underbrace{1\ldots1}_{o}\underbrace{0\ldots0}_{25-x-o})+1$ valid positions of the Os in the grid. Combining these two observations leads to Equation~\ref{eq:index} and guarantees that each state is associated to a unique natural number, and vice-versa.

In order to obtain fast computations, functions $pop(\cdot)$ and $ord(\cdot)$ are pre-computed and kept in memory at all time. Example values are presented in Table~\ref{table:pop_ord}. Overall, it takes less than a second to compute all such values, and it uses approximately 0.5GB of RAM.\footnote{There are $2^{25}$ input values $x$ and both $pop(x)$ and $ord(x)$ can be stored in a 32-bit variable. Hence, we obtain a total of $2^{25}\cdot32\cdot2=256$MB. It is also necessary to store the inverse function, namely the unique $x$ corresponding to a given pair $(pop(x),rd(c))$. For efficiency, the $x$ is stored in a 64-bit variable (to match the size of states), which makes a total of $2^{35}\cdot64=256$MB too.}

\begin{table}[h]
\centering
\caption{Population count $pop(x)$ and order $ord(x)$ of the first integers $x$}
\label{table:pop_ord}
\begin{tabular}{||c|c|c||c|c|c||}
$x$ & $pop(x)$ & $ord(x)$ & $x$ & $pop(x)$ & $ord(x)$\\
\hline
00000000 & 0 & 0 & 00001100 & 2 & 5\\
00000001 & 1 & 0 & 00001101 & 3 & 2\\ 
00000010 & 1 & 1 & 00001110 & 3 & 3\\ 
00000011 & 2 & 0 & 00001111 & 4 & 0\\ 
00000100 & 1 & 2 & 00010000 & 1 & 4\\
00000101 & 2 & 1 & 00010001 & 2 & 6\\
00000110 & 2 & 2 & 00010010 & 2 & 7\\
00000111 & 3 & 0 & 00010011 & 3 & 4\\
00001000 & 1 & 3 & 00010100 & 2 & 8\\
00001001 & 2 & 3 & 00010101 & 3 & 5\\
00001010 & 2 & 4 & 00010110 & 3 & 6\\
00001011 & 3 & 1 & 00010111 & 4 & 1\\
\end{tabular}
\end{table}

\section{Solving Quixo -- algorithms}\label{sec:algo}

\subsection{Basic algorithms}
\paragraph{Value iteration.}
After designing data structures, we now present our algorithms. As explained in Section~\ref{sec:goals}, due to cycles in the game ``tree'', minimax algorithm cannot be used for Quixo. The most natural algorithm for solving such games is the \emph{Value Iteration} (VI) algorithm (recalled in Algorithm~\ref{alg:value-iteration}). In the pseudo-code, the children of a state denote the set of states that are reachable after one move.\footnote{Some practical implementation details are omitted from pseudo-code. For example, since we only consider states with active player X, we need to swap the tiles/players after each move.} This algorithm follows closely the definition of outcome provided in Section~\ref{sec:prelim}:
\begin{itemize}
    \item Lines~1 to~7 deal with terminal states: a state is an immediate Win or Loss when there is a line of one of the symbols. Since we consider only positions where X is the active player, the existence of lines of Xs should be checked first. Indeed, by the rules, if there are lines of both symbols, the last player to move loses the game.
    \item Lines~8 to~14 iterate over the set of states until converging to a stable outcome. At the end of the computation, there may remain some Draw states.
\end{itemize}

\begin{algorithm}
\caption{Value Iteration (VI)}\label{alg:value-iteration}
\begin{algorithmic}[1]
\FORALL{states $s$}
    \IF{there is a line of Xs in $s$}
    \STATE $\text{outcome}[s]\leftarrow Win$
    \ELSIF{there is a line of Os in $s$}
    \STATE $\text{outcome}[s]\leftarrow Loss$
    \ELSE
    \STATE $\text{outcome}[s]\leftarrow Draw$
    \ENDIF
\ENDFOR

\REPEAT
    \FORALL{states $s$ such that $\text{outcome}[s] = Draw$}
    \IF{at least one child of $s$ is Loss}
    \STATE $\text{outcome}[s]\leftarrow Win$
    \ELSIF{all children of $s$ are Win}
    \STATE $\text{outcome}[s]\leftarrow Loss$
    \ENDIF
    \ENDFOR
\UNTIL{no update in the last iteration}
\end{algorithmic}
\end{algorithm}

\paragraph{Backward induction.}
Quixo cannot be solved directly applying this algorithm since it would require to store all outcomes at once in RAM (and thus would be too slow due to memory caching). Fortunately, we can use the classes ${\cal C}_{x,o}$ we defined in Section~\ref{sec:storage}.  Indeed, for any state of ${\cal C}_{x,o}$, its children belong to ${\cal C}_{o,x}\bigcup{\cal C}_{o,x+1}$ (due to player swap after each move). Thus, it becomes possible to compute all outcomes of ${\cal C}_{x,o}\bigcup{\cal C}_{o,x}$ using only ${\cal C}_{x,o+1}\bigcup{\cal C}_{o,x+1}$. Starting from states with 25 Xs or Os, and using \emph{backward induction}, we can compute all outcomes having only four classes of states in RAM at any given moment. The corresponding pseudo-code is given in Algorithm~\ref{alg:back-induction}.

\begin{algorithm}
\caption{Backward Induction using VI internally}\label{alg:back-induction}
\begin{algorithmic}[1]
\FOR{$n=25$ \TO 0}
    \FOR{$x=0$ \TO $\left\lceil n/2\right\rceil$}
        \STATE $o\leftarrow n-x$
        \IF{$n<25$}
        \STATE Load outcomes of classes ${\cal C}_{x,o+1}$ and ${\cal C}_{o,x+1}$
        \ENDIF
        \STATE Compute outcomes of classes ${\cal C}_{x,o}$ and ${\cal C}_{o,x}$ using VI
        \STATE Save outcomes of classes ${\cal C}_{x,o}$ and ${\cal C}_{o,x}$
        \STATE Unload all outcomes
    \ENDFOR
\ENDFOR
\end{algorithmic}
\end{algorithm}
This algorithm is likely to be able to solve Quixo, but, in practice, it is too slow to run on commodity hardware. Unfortunately, it is difficult to evaluate precisely its complexity because it depends on the number of internal (value) iterations, which is itself difficult to predict. The topology of the game ``tree'' has a strong impact on the required number of iterations to converge.

\subsection{Optimized algorithms}
We propose two algorithmic enhancements that significantly reduce the computation time.

\subsubsection{Use parent link}
To be a Win state, there should be at least one Loss child. Reversing this statement, we obtain that every parent of a Loss state is a Win state. This simple observation can be used to improve the computation. As soon as a state is found to be a Loss, we can compute all its parents and update their outcome to Win. Eventually they would have been updated to Win in Algorithm~\ref{alg:value-iteration}, but updating them immediately makes it possible to skip searching if states are Win (Lines 10 to 11) and may allow other states to be updated faster too. Note that parents of a given state can also be computed efficiently using a similar method as for computing its children (see Section~\ref{sec:state}). The pseudo-code is given in Algorithm~\ref{alg:update-from-term}. Note that the optimization is also used inside the internal iterations.

\begin{algorithm}
\caption{Update outcomes of ${\cal C}_{x,o}$ using terminal states}\label{alg:update-from-term}
\begin{algorithmic}[1]
\FORALL{states $s\in{\cal C}_{x,o}$}
    \IF{there is a line of Xs in $s$}
    \STATE $\text{outcome}[s]\leftarrow Win$
    \ELSIF{there is a line of Os in $s$}
    \STATE $\text{outcome}[s]\leftarrow Loss$
    \FORALL{parents $p\in {\cal C}_{o,x}$ of $s$}
        \STATE $\text{outcome}[p]\leftarrow Win$
    \ENDFOR
    \ENDIF
\ENDFOR
\end{algorithmic}
\end{algorithm}

\subsubsection{Use Win-or-Draw outcome}
Internal iterations require checking the outcomes of all children of a given state (see Lines~10 and~12 of Algorithm~\ref{alg:value-iteration}). Some of the children belong to already inductively-computed classes, while the others belong to classes currently being computed. More explicitly, for a state $s\in{\cal C}_{x,o}$, some children belong to ${\cal C}_{o,x}$ and some belong to ${\cal C}_{o,x+1}$. The outcome of this latter class has been already computed inductively. It is possible to check them only once by introducing a \emph{new temporary outcome}: WinOrDraw. For a given state $s\in{\cal C}_{x,o}$, among the children of $s$ in ${\cal C}_{o,x+1}$:
\begin{itemize}
    \item If there is at least one Loss state, then it is possible to immediately decide that $s$ is a Win state (as before).
    \item If all children are Win states, then the outcome of $s$ is initialized to Draw (as before). It means that it is still necessary to check children of $s$ in ${\cal C}_{o,x}$ to decide whether $s$ is a Win, Draw, or Loss.
    \item If there is at least one Draw state (and no Loss state), then the outcome of $s$ is initialized to WinOrDraw. As for the previous case, it is still necessary to check children of $s$ in ${\cal C}_{o,x}$ to decide the real outcome of $s$. However, we can already eliminate the Loss option since there is at least one Draw child.
\end{itemize}

The corresponding pseudo-code is given in Algorithm~\ref{alg:update-from-next}. Note that the test of Line~6 exists to avoid overriding a Win outcome already computed by Line~4.
\begin{algorithm}
\caption{Update outcomes of ${\cal C}_{x,o}$ using inductively-computed outcomes of ${\cal C}_{o,x+1}$ }\label{alg:update-from-next}
\begin{algorithmic}[1]
\FORALL{states $c\in {\cal C}_{o,x+1}$}
    \IF{$\text{outcome}[c] = Loss$}
    \FORALL{parents $s\in {\cal C}_{x,o}$ of $c$}
        \STATE $\text{outcome}[s]\leftarrow Win$
    \ENDFOR
    \ELSIF{$\text{outcome}[c] = Draw$}
    \FORALL{parents $s\in {\cal C}_{x,o}$ of $c$ s.t. $\text{outcome}[s]=Draw$}
        \STATE $\text{outcome}[s]\leftarrow WinOrDraw$
    \ENDFOR
    \ENDIF
\ENDFOR
\end{algorithmic}
\end{algorithm}

\subsubsection{Complete algorithm}
Combining value iteration, Backward induction, and our two optimizations, we obtain the complete algorithm we used to solve Quixo. It is summarized in Algorithm~\ref{alg:full-algo}. Using the additional WinOrDraw outcome allows for an additional subtle optimization. Inside internal iterations, the algorithm looks only for Loss states (\emph{i.e.}, states with only Win children). Win states are immediately updated when a child outcome is finalized as Loss. Therefore, it is not necessary to check if a state is Win. Since WinOrDraw states cannot become Loss states, there is no need to ever look at their children, hence the condition in Line~12.

\begin{algorithm*}
\caption{Complete algorithm used to solve Quixo $5\times5$.}\label{alg:full-algo}
\begin{algorithmic}[1]
\FOR{$n=25$ \TO 0}
    \FOR{$x=0$ \TO $\left\lceil n/2\right\rceil$}
        \STATE $o\leftarrow n-x$
        \FORALL{states $s\in {\cal C}_{x,o}\bigcup{\cal C}_{o,x}$}
            \STATE $\text{outcome}[s]\leftarrow Draw$
        \ENDFOR
        \STATE Update outcomes of ${\cal C}_{x,o}$ using terminal states (Algo~\ref{alg:update-from-term})
        \STATE Update outcomes of ${\cal C}_{o,x}$ using terminal states (Algo~\ref{alg:update-from-term})
        \IF{$n<25$}
        \STATE Update outcomes of ${\cal C}_{x,o}$ using inductively-computed outcomes of ${\cal C}_{o,x+1}$ (Algo~\ref{alg:update-from-next})
        \STATE Update outcomes of ${\cal C}_{o,x}$ using inductively-computed outcomes of ${\cal C}_{x,o+1}$ (Algo~\ref{alg:update-from-next})
        \ENDIF
        \REPEAT
            \FORALL{states $s\in {\cal C}_{x,o}\bigcup{\cal C}_{o,x}$ such that $\text{outcome}[s] = Draw$}
                \IF{all children $c\in {\cal C}_{x,o}\bigcup{\cal C}_{o,x}$ of $s$ are Win}
                \STATE $\text{outcome}[s]\leftarrow Loss$
                \FORALL{parents $p\in {\cal C}_{x,o}\bigcup{\cal C}_{o,x}$ of $s$}
                \STATE $\text{outcome}[p]\leftarrow Win$
                \ENDFOR
                \ENDIF
            \ENDFOR
        \UNTIL{no update in the last iteration}
        \FORALL{states $s\in {\cal C}_{x,o}\bigcup{\cal C}_{o,x}$ such that $\text{outcome}[s] = WinOrDraw$}
            \STATE $\text{outcome}[s]\leftarrow Draw$
        \ENDFOR
    \ENDFOR
\ENDFOR
\end{algorithmic}
\end{algorithm*}

\subsection{Parallelization}\label{sec:multithread}
Although Algorithm~\ref{alg:full-algo} is sequential, it is possible to run some of its parts in parallel in order to take advantage of multi-cores found in current commodity hardware. Let us first observe that previously computed outcomes can be accessed concurrently safely since they are only read by multiple threads. 

A high level parallelization is possible, by executing concurrently external iterations (Line~2 of Algorithm~\ref{alg:full-algo}). On the positive side, no synchronization is needed on the outcome variables (as they are written by different threads in different memory locations). On the negative side, though, significantly more memory is needed (each thread may need as much memory as the sequential algorithm we presented in Algorithm~\ref{alg:full-algo}). Furthermore, the number of states of ${\cal C}_{x,o}$ is large when $x=o$ or $x=o+1$. Now, since the other parts need to wait until the large states parts are computed, the speed improvement may be marginal and not scale linearly in the number of threads. Due to these concerns, we chose not to use this (simple) option. 

Our scheme for parallelization follows the rule: Replace each ``for all states \ldots do'' with a ``divide by $K$'' and execute $K$ threads in parallel. Of course, since threads now use the same memory locations for write access, we must use synchronisation mechanisms (\emph{e.g.} mutexes) to guarantee exclusive accesses to the currently-computed outcomes.\footnote{Implementation details: Due to our usage of \texttt{vector<bool>}, extra-caution was required. Accessing different elements of the same \texttt{vector<bool>} may lead to incorrect values: multiple bits are (typically) stored in the same byte, and only byte access can be guaranteed to be atomic by the processor.} 

\subsection{Deriving the optimal strategy}\label{sec:optimal}
Using Algorithm~\ref{alg:full-algo}, it is possible to compute all state outcomes. One may think that always choosing a Win action deterministically permits to win the game. Unfortunately, such a strategy does not guarantee winning since the Quixo game ``tree'' contains cycles. Hence, it is possible to enter a cycle where all states outcomes are Win, yet the game never finishes. Note that this behavior is unlike games that do not allow cycles in the game ``tree'' such as Connect-four.

It is possible to devise a probabilistically winning strategy by choosing a Win action uniformly at random: indeed, from any Win state, there exists a sequence of steps that does not belong to an infinite cycle. So, in an expected finite number of steps, the player wins. 

However, the random strategy is not necessarily optimal with respect to the number of steps taken to win. Instead, we focus on the strategy to win in the minimum number of steps (assuming the loosing player always chooses the action that delays her loss the most). Now, to actually compute the steps to win or lose, we now store the number of steps to the final outcome, using a new \emph{step} variable.
The \emph{step} variable is defined as follows:
\begin{itemize}
    \item In a terminal state, \emph{step} is $0$,
    \item If the state is Win, \emph{step} is one plus the minimum of the \emph{step}s of Loss children,
    \item If the state is Loss \emph{step} is one plus the maximum of the \emph{step}s of Win children.
\end{itemize}

Previous algorithms can be adapted to compute this additional \emph{step} variable. The Value Iteration algorithm changes to Algorithm~\ref{alg:vi-with-steps}.
\begin{algorithm}
\caption{Value Iteration with steps to Win or Loss}\label{alg:vi-with-steps}
\begin{algorithmic}[1]
\FORALL{states $s$}
    \IF{there is a line of Xs in $s$}
    \STATE $\text{outcome}[s]\leftarrow Win$
    \STATE $\text{step}[s]\leftarrow 0$
    \ELSIF{there is a line of Os in $s$}
    \STATE $\text{outcome}[s]\leftarrow Loss$
    \STATE $\text{step}[s]\leftarrow 0$
    \ELSE
    \STATE $\text{outcome}[s]\leftarrow Draw$
    \ENDIF
\ENDFOR

\STATE $i=0$
\REPEAT
    \FORALL{states $s$ such that $\text{outcome}[s] = Draw$}
    \IF{at least one child of $s$ exists whose outcome$[c]$ is Loss and step$[c]$ is $i-1$ }
    \STATE $\text{outcome}[s]\leftarrow Win$
    \STATE $\text{step}[s]\leftarrow i$
    \ELSIF{all children of $s$ are Win}
    \STATE $\text{outcome}[s]\leftarrow Loss$
    \STATE $\text{step}[s]\leftarrow \text{min(children steps)} + 1$
    \ENDIF
    \ENDFOR
    \STATE $i++$
\UNTIL{no update in last iteration}
\end{algorithmic}
\end{algorithm}

\newpage
\section{Results}\label{sec:result}
\subsection{$5\times5$ Quixo}
Our main result is that Quixo is a Draw game. In other words, if perfect players play the game, no one wins, that is, the game never finishes.

Using a single-thread computation,\footnote{We used a Ubuntu 18.04LTS server equipped with 32GB of RAM and powered by a 16-core Intel Core i9-9960X CPU.} it takes approximately $19\,500$ minutes (just under two weeks) to obtain this result. Using multithreading, as described in Section~\ref{sec:multithread}, the running time shrinks to around $1\,900$ minutes (i.e. $\approx32$ hours) using up to $K=32$ threads.

\begin{figure}[ht]
  \begin{center}
    \includegraphics[width=10.0cm]{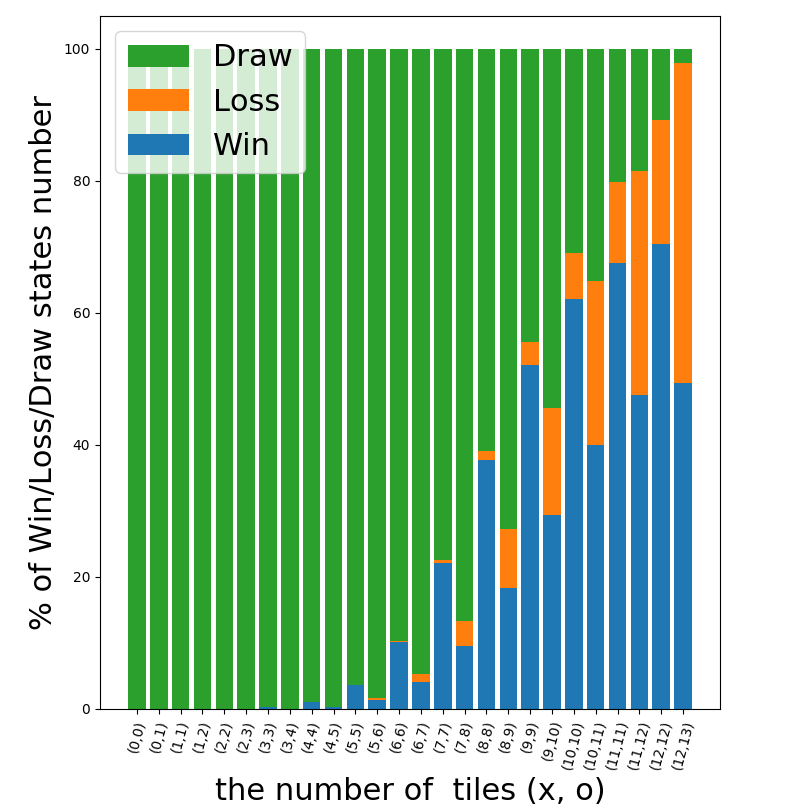}
    \caption{Percentage of Win, Loss, and Draw states for some classes ${\cal C}_{x,o}$. Player $X$ is next to play. The board size is $5\times5$}
    \label{fig:WLD-percentage}
  \end{center}
\end{figure}

\paragraph{Additional observations.}
From our computations, we were able to extract some interesting data that we present in the sequel. Table~\ref{table:total-WLD-number} shows the total number of Win, Loss, and Draw states. As the number of Draw states is much smaller than those of Win or Loss states, it may come to a surprise that the initial state is Draw. However, we also look at the distribution of these states. Most of the Draw states are located near the top of the game ``tree'' (\emph{i.e.}, with few marked tiles). Figure~\ref{fig:WLD-percentage} (when X is next to play) depicts the percentages of Win, Loss, and Draw states for some classes of states. We display classes where the number of $X$s and $O$s differ by at most 1. Intuitively, choosing the empty tile is a good strategy. The larger the number of marked tiles, the smaller the draw states percentage is. In other words, the latter part of the game is more complex. 

\begin{table}[ht]
\centering
\caption{Total Win, Loss and Draw states numbers}
\begin{tabular}{|c | c | c|} 
 \hline
 Win & Loss & Draw \\
 \hline
 441,815,157,309 & 279,746,227,956 & 125,727,224,178 \\
 \hline
\end{tabular}
\label{table:total-WLD-number}
\end{table}

\paragraph{Some interesting states.}
Figure~\ref{fig:quixo-x-win-5} shows an example where player $X$ can win (obviously, in more than one step). In the previous step, player $O$ chose a marked tile. Choosing a marked tile early in the game yields a big disadvantage. Moreover, the outcome of all reachable states in ${\cal C}_{2,1}$ with active player $X$ is $X$ wins. Figure~\ref{fig:quixo-x-win-same-tiles-5} shows another  example where player $X$ can win. Until this state, both players only chose empty tiles. Figure~\ref{fig:quixo-x-win-loses-same-tiles-5} shows an example where player $X$ loses. Such states are still complex for a human to understand the outcome.

\begin{figure*}[ht]
\begin{subfigure}[t]{0.3\textwidth}
    \centering
    \quixo[0.75]{5}{1,1,0,0,2, 0,0,0,0,0, 0,0,0,0,0, 0,0,0,0,0, 0,0,0,0,0}{}
    \caption{Player $X$ can win. This state is one of those with the smallest numbers of tiles such that the outcome is not draw. }
    \label{fig:quixo-x-win-5}
\end{subfigure}\hfill
\begin{subfigure}[t]{0.3\textwidth}
    \centering
    \quixo[0.75]{5}{1,1,0,1,0, 0,0,0,0,0, 0,0,0,0,0, 0,0,0,0,0, 0,2,0,2,2}{}
    \caption{Player $X$ can win. This state is one of those with the smallest numbers of tiles such that the outcome is not draw and both players have chosen empty tiles only.}
    \label{fig:quixo-x-win-same-tiles-5}
\end{subfigure}\hfill
\begin{subfigure}[t]{0.3\textwidth}
    \centering
    \quixo[0.75]{5}{2,2,2,2,0, 0,0,0,0,0, 0,0,0,1,0, 0,0,0,0,0, 0,0,1,1,1}{}
    \caption{Player $X$ loses. The number of Xs and Os tiles is the same but the active player loses.}
    \label{fig:quixo-x-win-loses-same-tiles-5}
\end{subfigure}\hfill
\caption{Some interesting states on the $5\times5$ board. Player $X$ is next to play.}
\end{figure*}

\subsection{$4\times4$ Quixo}
Contrarily to the real game, the $4\times4$ variant is a Win for the first player. Intuitively, the smaller board makes it easier to create a line. However, winning is not trivial; it requires up to $21$ moves when the opponent follows an optimal strategy. A complete optimal game is given in Figure~\ref{fig:4x4_optimal_play}.

\begin{figure}[h]
\centering
\input{animation_4x4_optimal.tex}  
\caption{Optimal play in $4\times4$ Quixo. The animation is unfortunately not working with all pdf readers. It works with Adobe Acrobat Reader. Non-animated play in Appendix~\ref{sec:appendix-optimal-play}}
\label{fig:4x4_optimal_play}
\end{figure}

\paragraph{Additional observations.}
Some states are obviously not reachable, e.g. a state containing a single $O$ not on an edge. Some other unreachable states are much less obvious, such as the state in Figure~\ref{fig:4x4_unreachable}. Globally, there are $41\,252\,106$ reachable states, which accounts for $95.8\%$ of the $3^{16}$ states. Therefore, ignoring unreachable states in the computation would not be significant.

\begin{figure*}[ht]
\begin{subfigure}[t]{0.2\textwidth}
    \centering
    \quixo[0.7]{4}{0,0,1,1, 2,2,2,1, 0,1,2,0, 1,0,2,0}{}
    \caption{Unreachable state. No previous state. }
\label{fig:4x4_unreachable}
\end{subfigure}\hfill
\begin{subfigure}[t]{0.2\textwidth}
    \centering
    \quixo[0.7]{4}{1,2,2,2, 2,2,1,2, 2,1,1,2, 2,2,2,1}{}
    \caption{Player $X$ loses in 1 step.}
    \label{fig:4x4_Lose_in_1step}
\end{subfigure}\hfill
\begin{subfigure}[t]{0.2\textwidth}
    \centering
    \quixo[0.7]{4}{0,0,2,1, 0,0,0,0, 1,2,2,0, 0,0,0,0}{}
    \caption{Player $X$ loses in 22 steps.}
    \label{fig:4x4_Lose_in_22step}
\end{subfigure}\hfill
\begin{subfigure}[t]{0.3\textwidth}
    \centering
    \quixo[0.7]{4}{0,1,1,0, 2,0,2,1, 2,1,1,1, 2,1,1,2}{} 
    \caption{Draw state. $O$ can come back to this state (or a symmetric one) with the next $O$ step even if $X$ plays optimally.}
    \label{fig:4x4_draw_o8x5}
\end{subfigure}
\caption{Some interesting states on the $4\times4$ board. Player $X$ is next to play.}
\end{figure*}

Using the algorithm described in Section~\ref{sec:optimal},
Table~\ref{table:win-loss-step} shows the number of steps to Win and Loss for optimal strategies. The winner selects an action that minimizes the number of steps to a Win, and the loser selects an action that maximizes the number of steps to a Loss. 

Normally, a winner wins in an odd number of steps, and a loser loses in an even number of steps. However, some states yield an optimal player to lose in $1$ step. One such example is shown in Figure~\ref{fig:4x4_Lose_in_1step}. In all next states, there is a line of $O$, so $X$ loses in $1$ step. 

Another interesting result is that there are some states that lose in $22$ steps although no state wins in $23$ steps, and the initial state wins in $21$ steps. Figure~\ref{fig:4x4_Lose_in_22step} is the example of a state to lose in $22$ steps; only its symmetric states are the those that lose in 22 steps. 

\begin{table}[h!]
\centering
\caption{Steps to Win and Loss of an optimal strategy}
\begin{tabular}{|ccc||ccc|} 
 \hline
 steps & Win states & Loss states & steps & Win states & Loss states \\
 \hline
0 & 4,697,505 & 4,530,779 & 12 & 0 & 182,954 \\
1 & 15,277,446 & 528 & 13 & 100,374 & 0 \\
2 & 0 & 3,775,611 & 14 & 0 & 66,280 \\
3 & 2,419,938 & 0 & 15 & 29,314 & 0 \\
4 & 0 & 2,970,384 & 16 & 0 & 18,014 \\
5 & 1,740,992 & 0 & 17 & 6,656 & 0  \\
6 & 0 & 1,982,339 & 18 & 0 & 4,084    \\
7 & 1,214,497 & 0 & 19 & 1,012 & 0    \\
8 & 0 & 1,034,097 & 20 & 0 & 520      \\
9 & 658,834 & 0 & 21 & 57 & 0       \\ 
10 & 0 & 438,138 & 22 & 0 & 8        \\
11 & 287,864 & 0 & 23 & 0 & 0        \\
 \hline
 &&&total & 26,434,489 & 15,003,736 \\
 \hline
\end{tabular}
\label{table:win-loss-step}
\end{table}



\section{Conclusions and open questions}\label{sec:conclusion}

To summarize, the official $5\times5$ Quixo is a Draw game; neither player can win. Smaller $3\times3$ and $4\times4$ variants are First-Player-Win games.\footnote{The $3\times3$ version is not discussed in this paper and is left to the reader. The first player wins in 7~moves.}

Given that the $5\times5$ board is already a Draw game, one may expect larger instances to be Draw games too. We conjecture that it is the case, but we were not able to prove it.

In a different direction, one may be interested in the complexity of Quixo. 
Mishiba and Takenaga already studied the complexity of a generalization of Quixo~\cite{quixo-complexity}. They proved the game to be EXPTIME-complete. In the paper, they consider arbitrary large boards, but players still have to align only five identical symbols. Keeping the required line length equal to the board size may change the complexity.

Based on the previous generalization, a natural question arises; can the first player (or unlikely the second player) create a line of four symbols when playing on the $5\times5$ board? Changing the two lines losing rule into a winning rule may also change the global outcome.
Finally, a last research direction would be to compute human-playable optimal strategies. We strongly solved Quixo on $4\times 4$ and $5\times 5$ grids. However, playing an optimal strategy remains difficult for humans. Nim is mathematically solved~\cite{bouton1901nim}, and following an optimal strategy is not so difficult for humans. Finding an optimal strategy for Quixo that can be remembered by humans is still an open problem.

\bibliographystyle{unsrt}
\bibliography{ms}

\clearpage
\appendix
\section{Inanimate version of Figure~\ref{fig:4x4_optimal_play}}
\label{sec:appendix-optimal-play}
\newcommand\inanimatedQuixo[4][1]{
	\begin{tikzpicture}[y=-1cm,scale=#1,every node/.style={scale=#1}]
	\draw[draw=none] (-1,-1.5) rectangle (#2,#2+0.5);
	\ifthenelse{\equal{#4}{}}{}{\node at ({(#2-1)/2},#2) {\LARGE Move #4};}
	\draw[ultra thick,rounded corners] (-0.5,-0.5) rectangle (#2-0.5,#2-0.5);
	\foreach \x [count=\xi] in {#3} {
		\tikzmath{\a = {int(\x/10)};   
						\b = {int(Mod(\x, 10))};   
						if \b == 0 then {\dx = 0; \dy=0;} else {
						if \b == 1 then {\dx = -0.5; \dy=0;} else {
						if \b == 2 then {\dx = 0.5; \dy=0;} else {
						if \b == 3 then {\dx = 0; \dy=-0.5;} else {
						if \b == 4 then {\dx = 0; \dy=0.5;} else {
						if \b == 5 then {\dx = -1; \dy=0;} else {
						if \b == 6 then {\dx = 1; \dy=0;} else {
						if \b == 7 then {\dx = 0; \dy=-1;} else {
						if \b == 8 then {\dx = 0; \dy=1;} else {
						};};};};};};};};};
		}
		\ifthenelse{\a = 1}{\node[draw, rectangle, rounded corners, fill=black!5, minimum size = 0.98cm] at ({mod(\xi-1,#2)+\dx},{int((\xi-1)/#2)+\dy}) {
		};}{
		\ifthenelse{\a = 2}{\node[draw, rectangle, rounded corners, fill=black!5, minimum size = 0.98cm] at ({mod(\xi-1,#2)+\dx},{int((\xi-1)/#2)+\dy}) {
			\begin{tikzpicture}\draw[thick] (-0.25,0.25) -- (0.25,-0.25) (-0.25,-0.25) -- (0.25,0.25);\end{tikzpicture}
		};}{
		\ifthenelse{\a = 3}{\node[draw, rectangle, rounded corners, fill=black!5, minimum size = 0.98cm, anchor=center] at ({mod(\xi-1,#2)+\dx},{int((\xi-1)/#2)+\dy}) {
			\begin{tikzpicture}\draw[thick] (0,0) circle (0.25);\end{tikzpicture}
		};}{
		\ifthenelse{\a = 4}{\node[draw, rectangle, rounded corners, fill=red!25, minimum size = 0.98cm] at ({mod(\xi-1,#2)+\dx},{int((\xi-1)/#2)+\dy}) {
		};}{
		\ifthenelse{\a = 5}{\node[draw, rectangle, rounded corners, fill=red!25, minimum size = 0.98cm] at ({mod(\xi-1,#2)+\dx},{int((\xi-1)/#2)+\dy}) {
			\begin{tikzpicture}\draw[thick] (-0.25,0.25) -- (0.25,-0.25) (-0.25,-0.25) -- (0.25,0.25);\end{tikzpicture}
		};}{
		\ifthenelse{\a = 6}{\node[draw, rectangle, rounded corners, fill=blue!25, minimum size = 0.98cm, anchor=center] at ({mod(\xi-1,#2)+\dx},{int((\xi-1)/#2)+\dy}) {
			\begin{tikzpicture}\draw[thick] (0,0) circle (0.25);\end{tikzpicture}
		};}{
		\ifthenelse{\a = 7}{\node[draw, rectangle, rounded corners, fill=blue!25, minimum size = 0.98cm] at ({mod(\xi-1,#2)+\dx},{int((\xi-1)/#2)+\dy}) {
		};};};};};};};}
	}
	\end{tikzpicture}
}

\begin{center}
\inanimatedQuixo[0.7]{4}{10,10,10,10, 10,10,10,10, 10,10,10,10, 10,10,10,10}{}
\inanimatedQuixo[0.7]{4}{10,10,10,10, 40,10,10,10, 10,10,10,10, 10,10,10,10}{1}
\inanimatedQuixo[0.7]{4}{10,10,10,10, 12,12,12,52, 10,10,10,10, 10,10,10,10}{1}
\inanimatedQuixo[0.7]{4}{10,10,10,10, 10,10,10,50, 10,10,10,10, 10,10,10,10}{1}

\inanimatedQuixo[0.7]{4}{10,10,10,10, 10,10,10,20, 10,10,10,10, 10,10,10,10}{}
\inanimatedQuixo[0.7]{4}{70,10,10,10, 10,10,10,20, 10,10,10,10, 10,10,10,10}{2}
\inanimatedQuixo[0.7]{4}{14,10,10,10, 14,10,10,20, 14,10,10,10, 64,10,10,10}{2}
\inanimatedQuixo[0.7]{4}{10,10,10,10, 10,10,10,20, 10,10,10,10, 60,10,10,10}{2}

\inanimatedQuixo[0.7]{4}{10,10,10,10, 10,10,10,20, 10,10,10,10, 30,10,10,10}{}
\inanimatedQuixo[0.7]{4}{10,10,10,10, 10,10,10,20, 10,10,10,10, 30,10,10,40}{3}
\inanimatedQuixo[0.7]{4}{10,10,10,10, 10,10,10,20, 10,10,10,10, 51,31,11,11}{3}
\inanimatedQuixo[0.7]{4}{10,10,10,10, 10,10,10,20, 10,10,10,10, 50,30,10,10}{3}

\inanimatedQuixo[0.7]{4}{10,10,10,10, 10,10,10,20, 10,10,10,10, 20,30,10,10}{}
\inanimatedQuixo[0.7]{4}{10,10,10,70, 10,10,10,20, 10,10,10,10, 20,30,10,10}{4}
\inanimatedQuixo[0.7]{4}{10,10,10,24, 10,10,10,14, 10,10,10,14, 20,30,10,64}{4}
\inanimatedQuixo[0.7]{4}{10,10,10,20, 10,10,10,10, 10,10,10,10, 20,30,10,60}{4}

\inanimatedQuixo[0.7]{4}{10,10,10,20, 10,10,10,10, 10,10,10,10, 20,30,10,30}{}
\inanimatedQuixo[0.7]{4}{10,10,10,20, 10,10,10,40, 10,10,10,10, 20,30,10,30}{5}
\inanimatedQuixo[0.7]{4}{10,10,10,20, 10,10,10,14, 10,10,10,34, 20,30,10,54}{5}
\inanimatedQuixo[0.7]{4}{10,10,10,20, 10,10,10,10, 10,10,10,30, 20,30,10,50}{5}

\inanimatedQuixo[0.7]{4}{10,10,10,20, 10,10,10,10, 10,10,10,30, 20,30,10,20}{}
\inanimatedQuixo[0.7]{4}{70,10,10,20, 10,10,10,10, 10,10,10,30, 20,30,10,20}{6}
\inanimatedQuixo[0.7]{4}{14,10,10,20, 14,10,10,10, 24,10,10,30, 64,30,10,20}{6}
\inanimatedQuixo[0.7]{4}{10,10,10,20, 10,10,10,10, 20,10,10,30, 60,30,10,20}{6}

\inanimatedQuixo[0.7]{4}{10,10,10,20, 10,10,10,10, 20,10,10,30, 30,30,10,20}{}
\inanimatedQuixo[0.7]{4}{40,10,10,20, 10,10,10,10, 20,10,10,30, 30,30,10,20}{7}
\inanimatedQuixo[0.7]{4}{14,10,10,20, 24,10,10,10, 34,10,10,30, 54,30,10,20}{7}
\inanimatedQuixo[0.7]{4}{10,10,10,20, 20,10,10,10, 30,10,10,30, 50,30,10,20}{7}

\inanimatedQuixo[0.7]{4}{10,10,10,20, 20,10,10,10, 30,10,10,30, 20,30,10,20}{}
\inanimatedQuixo[0.7]{4}{10,10,10,20, 20,10,10,10, 30,10,10,30, 20,30,70,20}{8}
\inanimatedQuixo[0.7]{4}{10,10,10,20, 20,10,10,10, 30,10,10,30, 20,30,22,62}{8}
\inanimatedQuixo[0.7]{4}{10,10,10,20, 20,10,10,10, 30,10,10,30, 20,30,20,60}{8}

\inanimatedQuixo[0.7]{4}{10,10,10,20, 20,10,10,10, 30,10,10,30, 20,30,20,30}{}
\inanimatedQuixo[0.7]{4}{10,10,10,20, 20,10,10,40, 30,10,10,30, 20,30,20,30}{9}
\inanimatedQuixo[0.7]{4}{10,10,10,20, 20,10,10,34, 30,10,10,34, 20,30,20,54}{9}
\inanimatedQuixo[0.7]{4}{10,10,10,20, 20,10,10,30, 30,10,10,30, 20,30,20,50}{9}

\inanimatedQuixo[0.7]{4}{10,10,10,20, 20,10,10,30, 30,10,10,30, 20,30,20,20}{}
\inanimatedQuixo[0.7]{4}{10,10,70,20, 20,10,10,30, 30,10,10,30, 20,30,20,20}{10}
\inanimatedQuixo[0.7]{4}{10,10,14,20, 20,10,14,30, 30,10,24,30, 20,30,64,20}{10}
\inanimatedQuixo[0.7]{4}{10,10,10,20, 20,10,10,30, 30,10,20,30, 20,30,60,20}{10}

\inanimatedQuixo[0.7]{4}{10,10,10,20, 20,10,10,30, 30,10,20,30, 20,30,30,20}{}
\inanimatedQuixo[0.7]{4}{40,10,10,20, 20,10,10,30, 30,10,20,30, 20,30,30,20}{11}
\inanimatedQuixo[0.7]{4}{12,12,22,52, 20,10,10,30, 30,10,20,30, 20,30,30,20}{11}
\inanimatedQuixo[0.7]{4}{10,10,20,50, 20,10,10,30, 30,10,20,30, 20,30,30,20}{11}

\inanimatedQuixo[0.7]{4}{10,10,20,20, 20,10,10,30, 30,10,20,30, 20,30,30,20}{}
\inanimatedQuixo[0.7]{4}{10,70,20,20, 20,10,10,30, 30,10,20,30, 20,30,30,20}{12}
\inanimatedQuixo[0.7]{4}{61,11,20,20, 20,10,10,30, 30,10,20,30, 20,30,30,20}{12}
\inanimatedQuixo[0.7]{4}{60,10,20,20, 20,10,10,30, 30,10,20,30, 20,30,30,20}{12}

\inanimatedQuixo[0.7]{4}{30,10,20,20, 20,10,10,30, 30,10,20,30, 20,30,30,20}{}
\inanimatedQuixo[0.7]{4}{30,40,20,20, 20,10,10,30, 30,10,20,30, 20,30,30,20}{13}
\inanimatedQuixo[0.7]{4}{51,31,20,20, 20,10,10,30, 30,10,20,30, 20,30,30,20}{13}
\inanimatedQuixo[0.7]{4}{50,30,20,20, 20,10,10,30, 30,10,20,30, 20,30,30,20}{13}

\inanimatedQuixo[0.7]{4}{20,30,20,20, 20,10,10,30, 30,10,20,30, 20,30,30,20}{}
\inanimatedQuixo[0.7]{4}{20,30,20,20, 20,10,10,30, 30,10,20,30, 20,30,60,20}{14}
\inanimatedQuixo[0.7]{4}{20,30,63,20, 20,10,23,30, 30,10,13,30, 20,30,23,20}{14}
\inanimatedQuixo[0.7]{4}{20,30,60,20, 20,10,20,30, 30,10,10,30, 20,30,20,20}{14}

\inanimatedQuixo[0.7]{4}{20,30,30,20, 20,10,20,30, 30,10,10,30, 20,30,20,20}{}
\inanimatedQuixo[0.7]{4}{20,30,30,20, 20,10,20,30, 30,10,10,30, 50,30,20,20}{15}
\inanimatedQuixo[0.7]{4}{20,30,30,20, 20,10,20,30, 30,10,10,30, 32,22,22,52}{15}
\inanimatedQuixo[0.7]{4}{20,30,30,20, 20,10,20,30, 30,10,10,30, 30,20,20,50}{15}

\inanimatedQuixo[0.7]{4}{20,30,30,20, 20,10,20,30, 30,10,10,30, 30,20,20,20}{}
\inanimatedQuixo[0.7]{4}{20,30,30,20, 20,10,20,30, 30,10,10,60, 30,20,20,20}{16}
\inanimatedQuixo[0.7]{4}{20,30,30,20, 20,10,20,30, 30,10,10,24, 30,20,20,64}{16}
\inanimatedQuixo[0.7]{4}{20,30,30,20, 20,10,20,30, 30,10,10,20, 30,20,20,60}{16}

\inanimatedQuixo[0.7]{4}{20,30,30,20, 20,10,20,30, 30,10,10,20, 30,20,20,30}{}
\inanimatedQuixo[0.7]{4}{20,30,30,20, 20,10,20,30, 30,10,10,50, 30,20,20,30}{17}
\inanimatedQuixo[0.7]{4}{20,30,30,20, 20,10,20,30, 51,31,11,11, 30,20,20,30}{17}
\inanimatedQuixo[0.7]{4}{20,30,30,20, 20,10,20,30, 50,30,10,10, 30,20,20,30}{17}

\inanimatedQuixo[0.7]{4}{20,30,30,20, 20,10,20,30, 20,30,10,10, 30,20,20,30}{}
\inanimatedQuixo[0.7]{4}{20,30,30,20, 20,10,20,60, 20,30,10,10, 30,20,20,30}{18}
\inanimatedQuixo[0.7]{4}{20,30,30,20, 61,21,11,21, 20,30,10,10, 30,20,20,30}{18}
\inanimatedQuixo[0.7]{4}{20,30,30,20, 60,20,10,20, 20,30,10,10, 30,20,20,30}{18}

\inanimatedQuixo[0.7]{4}{20,30,30,20, 30,20,10,20, 20,30,10,10, 30,20,20,30}{}
\inanimatedQuixo[0.7]{4}{20,30,30,20, 30,20,10,20, 20,30,10,40, 30,20,20,30}{19}
\inanimatedQuixo[0.7]{4}{20,30,30,20, 30,20,10,20, 20,30,10,34, 30,20,20,54}{19}
\inanimatedQuixo[0.7]{4}{20,30,30,20, 30,20,10,20, 20,30,10,30, 30,20,20,50}{19}

\inanimatedQuixo[0.7]{4}{20,30,30,20, 30,20,10,20, 20,30,10,30, 30,20,20,20}{}
\inanimatedQuixo[0.7]{4}{20,30,30,20, 60,20,10,20, 20,30,10,30, 30,20,20,20}{20}
\inanimatedQuixo[0.7]{4}{20,30,30,20, 22,12,22,62, 20,30,10,30, 30,20,20,20}{20}
\inanimatedQuixo[0.7]{4}{20,30,30,20, 20,10,20,60, 20,30,10,30, 30,20,20,20}{20}

\inanimatedQuixo[0.7]{4}{20,30,30,20, 20,10,20,30, 20,30,10,30, 30,20,20,20}{}
\inanimatedQuixo[0.7]{4}{20,30,30,20, 20,10,20,30, 50,30,10,30, 30,20,20,20}{21}
\inanimatedQuixo[0.7]{4}{20,30,30,20, 20,10,20,30, 34,30,10,30, 54,20,20,20}{21}
\inanimatedQuixo[0.7]{4}{20,30,30,20, 20,10,20,30, 30,30,10,30, 50,20,20,20}{21}

\inanimatedQuixo[0.7]{4}{20,30,30,20, 20,10,20,30, 30,30,10,30, 20,20,20,20}{}
\inanimatedQuixo[0.7]{4}{20,30,30,20, 20,10,20,30, 30,30,10,30, 50,50,50,50}{}
\inanimatedQuixo[0.7]{4}{20,30,30,20, 20,10,20,30, 30,30,10,30, 20,20,20,20}{}
\inanimatedQuixo[0.7]{4}{20,30,30,20, 20,10,20,30, 30,30,10,30, 50,50,50,50}{}
\end{center}  

\end{document}